\documentclass[aps,prl,reprint,floatfix,a4paper]{revtex4-1}  

\pdfoutput=1
\usepackage[T1]{fontenc}
\usepackage[utf8x]{inputenc}

\usepackage{mathptmx}
\usepackage[Symbol]{upgreek}
\usepackage{times}
\usepackage{fixmath}

\usepackage{helvet}

\usepackage{hyperref}

\usepackage{xcolor}
\usepackage{graphicx}
\usepackage{amsmath,amssymb}

\usepackage{geometry}
\pdfpageattr {/Group << /S /Transparency /I true /CS /DeviceRGB>>}

\usepackage{setspace}

\usepackage[stretch=15,shrink=15,step=3]{microtype}
\usepackage{cmap}

\hypersetup{pdftex, bookmarks, pdffitwindow=false, pdfstartview=FitH, pdfdisplaydoctitle, colorlinks, plainpages=false, pdfpagelabels, hypertexnames, citecolor={blue},linkcolor={red}, urlcolor={violet}, pdflang={en}, hyperfootnotes=false, breaklinks, pdftitle={Computing phase diagrams for a quasicrystal-forming patchy-particle system}}

\newcommand{\pdc}[3]{\ensuremath{\left( \frac{\partial #1}{\partial #2}\right)_{#3}}}
\newcommand{\der}{\ensuremath{\mathrm{d}}}
\def\avg#1{\ensuremath{\left\langle #1 \right\rangle}}
\newcommand{\deriv}[2]{\ensuremath{\frac{\mathrm{d}#1}{\mathrm{d}#2}}}

\begin{document}

\title{Computing phase diagrams for a quasicrystal-forming patchy-particle
system}
\author{Aleks Reinhardt}
\author{Flavio Romano}
\author{Jonathan P.~K.~Doye}
\email[Correspondence author. E-mail: ]{jonathan.doye@chem.ox.ac.uk}
\affiliation{Physical and Theoretical Chemistry Laboratory, Department of
Chemistry, University of Oxford, Oxford, OX1 3QZ, United Kingdom}
\date{8 April 2013}

\begin{abstract}
We introduce an approach to computing the free energy of quasicrystals, which we
use to calculate phase diagrams for systems of two-dimensional patchy particles
with five regularly arranged patches that have previously been shown to form
dodecagonal quasicrystals. We find that the quasicrystal is a thermodynamically
stable phase for a wide range of conditions and remains a robust feature of the
system as the potential's parameters are varied. We also demonstrate that the
quasicrystal is entropically stabilised over its crystalline approximants.
\end{abstract}



\maketitle

Quasicrystals are a type of aperiodic crystal structure: they have well-ordered
structures, but cannot be described by a periodic lattice, not even one with
incommensurate lattice parameters \cite{Steurer2004, *Steurer2012}. They were
first reported by Shechtman \textit{et al.}\ \cite{Shechtman1984}, in whose work
a rapidly cooled Al-Mn alloy was found to result in a metastable quasicrystal.
Most quasicrystals discovered so far are metallic alloys
\cite{Steurer2004, *Steurer2012}; however, recently, an increasing number of
examples have been reported in the field of soft condensed matter
\cite{Zeng2005, *Fischer2011, *Dotera2011}. Quasicrystals have also been
observed to form in computer simulations, in which not only binary, but also
one-component quasicrystals have been seen \cite{Widom1987, *Leung1989,
*Dzugutov1993, *Skibinsky1999, *Engel2007, *Keys2007, *Johnston2010b,
*Johnston2010c, *ZhiWei2012, *Dotera2012,  HajiAkbari2009, HajiAkbari2011,
*HajiAkbari2011b, Iacovella2011, VanDerLinden2012}. While certain quasicrystals
have been shown to be stable in experiment \cite{Tsai2003}, in simulations, the
relatively short timescales accessible mean that it is not necessarily clear
whether quasicrystals are truly the stable phase, or rather a metastable phase
that is more kinetically accessible. In this work, we address this important
question of assessing the stability of quasicrystals and apply it to a soft
matter system that has the potential to be realised experimentally.

To prove that quasicrystals are thermodynamically stable, their free energy must
be computed. However, such a calculation is not straightforward: the principal
problem is that there is no obvious reference state whose free energy is
known and from which thermodynamic integration \cite{Vega2008, Frenkel1984}
could be used to calculate the free energy of the quasicrystalline phase. A
phase transition intervenes when integrating from an ideal gas, which is
normally used as a reference state for fluid phases, and an Einstein crystal,
used as a reference state for crystalline phases, would also not be suitable
because it would fail to capture the configurational entropy associated with the
quasicrystal's many possible structures.

One approach to assessing quasicrystal stability is to compute the free energy
of an approximant crystalline phase. However, it is not immediately obvious
whether quasicrystals are more or less stable than their approximants
\cite{HajiAkbari2009, HajiAkbari2011, *HajiAkbari2011b};  whilst the enthalpy
and vibrational properties of the quasicrystal and its approximant phases are
likely to be similar, the free energy of a quasicrystal has a contribution from
its configurational entropy that is not present for the approximant
\cite{Iacovella2011}. Nevertheless, such calculations have provided some insight
into the phase behaviour of systems such as hard tetrahedra and bipyramids
\cite{HajiAkbari2011, *HajiAkbari2011b} and spherical micelles
\cite{Iacovella2011}. By contrast, Kiselev \textit{et al.}\ recently estimated
the free energy of the quasicrystal itself by combining a phonon contribution
for a particular quasicrystal configuration from thermodynamic integration and a
configurational contribution based on an approximation of uncorrelated phason
flips \cite{Kiselev2012}. They confirmed that, within their approximation, the
quasicrystal's free energy is in certain conditions lower than the
approximant's.

Our solution to this conundrum is to note that thermodynamic integration is not
the only way the melting point of a solid can be computed. Another method is to
simulate the direct coexistence of two phases with an interface \cite{Vega2008}.
By performing such simulations at a range of temperatures, we can bracket the
regions where the quasicrystal melts ($T>T_\text{fus}$) and  grows
($T<T_\text{fus}$). At $T=T_\text{fus}$, the chemical potential of the
quasicrystal is equal to that of the fluid, which we can calculate using
thermodynamic integration. Such an equilibrium approach by construction accounts
for the quasicrystal's configurational entropy. Once the quasicrystal's free
energy is known at one point, we can use thermodynamic integration
to reach other points of interest on the phase diagram.

Here, we apply this approach to a two-dimensional system of patchy particles
\cite{Glotzer2007,*Pawar2010} with five regularly arranged attractive patches,
which we model using a simple potential that has previously been used in
simulations of self-assembly and crystallisation \cite{Wilber2007, *Noya2007,
*Noya2010, *Williamson2011b, *Doppelbauer2012b, Doye2007, VanDerLinden2012,
Doppelbauer2010}. The potential is based on the Lennard-Jones (LJ) form,
$V^{\text{LJ}}(r_{ij}) = 4 \varepsilon [(\sigma_\text{LJ}/r_{ij})^{2n}-
(\sigma_\text{LJ}/r_{ij} )^{n}]$, where $n=6$ for the standard LJ potential,
where for $r_{ij}>\sigma_\text{LJ}$, this potential is multiplied by an angular
modulation factor \begin{equation} V^{\text{A}} =
\exp\left[\frac{-\theta_{k_\text{min}ij}^2}{2\,\sigma_\text{pw}^2}\right]
\exp\left[\frac{-\theta_{l_\text{min}ji}^2}{2\,\sigma_\text{pw}^2}\right],
\end{equation} where $\sigma_\text{pw}$ is the `patch width' (in radians) and
$\theta_{kij}$ is the angle between the patch vector of patch $k$ (chosen to
minimise this angle) on particle $i$ and the interparticle vector
$\mathbold{r}_{ij}$.

This interaction potential provides a simple model for the patchy colloidal
particles that many experimental groups have, with progressively increasing
success, been seeking to develop \cite{DeVries2007, *Cho2007, *Yang2008,
*Kraft2009,  *Wang2008, *Mao2010, *Duguet2011, *Wang2012, Chen2011,
Glotzer2007,*Pawar2010}. The enhanced range of structural behaviour that such
patchy interactions could facilitate has been extensively studied in computer
simulation \cite{Wilber2007, *Noya2007,  *Noya2010,  *Williamson2011b,
*Doppelbauer2012b, Doye2007, VanDerLinden2012, Doppelbauer2010, Kern2003,
*Zhang2004, *Bianchi2006, *Sciortino2009, *Romano2011b, *Bianchi2011}.

\begin{figure}
\begin{center}
\includegraphics{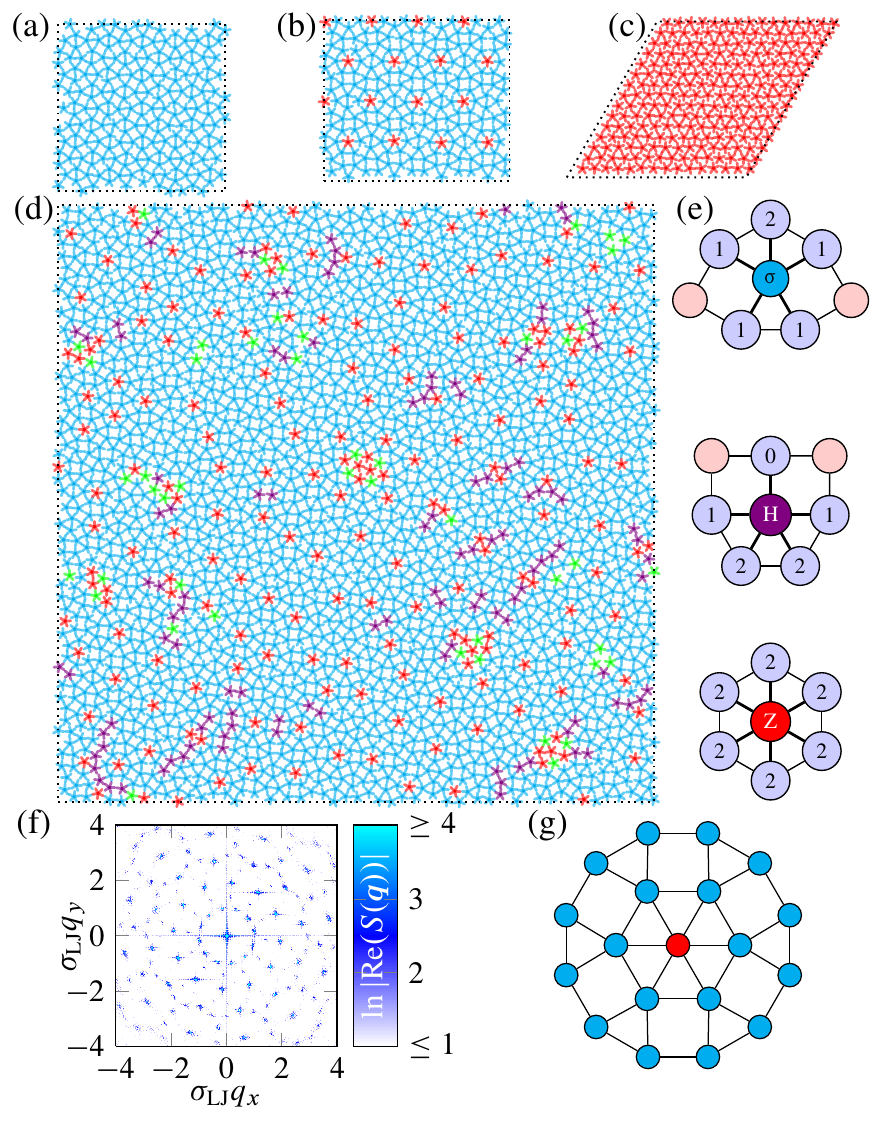}
\end{center}
\caption{Examples of the main phases studied. (a) $\upsigma$
phase. (b) Approximant crystal.  (c) Plastic hexagonal (Z) phase. (d)
Quasicrystal, $N=2500$. (e) Neighbour classification of $\upsigma$, H and Z
environments. The numbers of common neighbours are given for each of the central
particle's neighbours. These local environments are used to colour the particles
in (a)--(d), where in addition, particles that do not adopt one of these
environments are coloured green. (e) Diffraction pattern corresponding to the
configuration shown in (a). (f) Dodecagonal motif characteristic of the
quasicrystal.}\label{fig-quasicrystal}
\end{figure}

Dodecagonal quasicrystals were previously found to form on cooling systems of
these five-patch particles in certain conditions, and structures with the lowest
enthalpy were also identified \cite{VanDerLinden2012}. Some of the relevant
phases studied are shown in Fig.~\ref{fig-quasicrystal}(a--d); particles are
classified based on a common neighbour analysis \cite{VanDerLinden2012} into
$\upsigma$, H and Z environments (shown in Fig.~\ref{fig-quasicrystal}(e)), by
analogy to the Frank--Kasper phases \cite{Frank1958, *Frank1959}. The hexagonal
(Z) phase (Fig.~\ref{fig-quasicrystal}(c)) was found to form at high pressures
(as it is densest) and for wide patches (where the potential is closer to
being isotropic). At low pressures and at reasonably narrow patch widths, the
structure best satisfying the attractive patches is the enthalpically
favoured phase. A crystal cannot exist with five-fold symmetry, and so no
crystal with five perfectly aligned patches exists. As a compromise, each
particle has five neighbours, but the angles between the neighbours do not
perfectly match the five-fold symmetry of the particle itself. Local
environments satisfying this requirement are the $\upsigma$ and H environments
shown in Fig.~\ref{fig-quasicrystal}(e); the lowest enthalpy structure at low
pressure and reasonably narrow patch widths is the $\upsigma$ crystal shown in
Fig.~\ref{fig-quasicrystal}(a) \cite{VanDerLinden2012}.

Quasicrystals were observed in some cooling runs in the region where the
$\upsigma$ phase is the lowest in enthalpy, but
near to the hexagonal `boundary'. The structure of a typical quasicrystal is
shown in Fig.~\ref{fig-quasicrystal}(d); its diffraction pattern
(Fig.~\ref{fig-quasicrystal}(f)) exhibits a characteristic twelve-fold symmetry.
The local environments in the quasicrystal are predominantly of the $\upsigma$
type, but a significant fraction of Z environments can be seen: these are
typically located at the centre of a dodecagonal motif
(Fig.~\ref{fig-quasicrystal}(g)). Much of the structure can be analysed in terms
of packing of such dodecagons into triangular, square and rectangular
arrangements \cite{VanDerLinden2012}, which can readily be seen in
Fig.~\ref{fig-quasicrystal}(d); the various possible ways of arranging these
dodecagons are likely to lead to substantial entropy. There is no translational
order, but bonds can be orientated in any one of twelve directions, resulting in
long-range orientational order of dodecagonal character.

We perform Monte Carlo simulations in the $NpT$ ensemble using a
rectangular box with periodic boundaries \footnote{See supplementary material in the appendix for implementation details and a summary of
simulation methods used.}. As a starting point in the determination of phase
diagrams for this system, we chose a patch width, temperature and pressure at
which the quasicrystal is known to form on cooling, and thus locate the
fluid-quasicrystal (F-QC) equilibrium transition. This transition is mostly
rapid and facile, with essentially no hysteresis, which allows us to determine
the coexistence points directly by performing simulations in relatively large
boxes starting from both phases. Whilst we initially performed direct
coexistence simulations to determine the melting point of the quasicrystal,
these simulations mostly did not prove to be any more efficient than direct
brute-force simulations, in which no initial interface was introduced, as the
growth of the phases was not restricted to the initial interface.
This lack of hysteresis is most likely indicative of a particularly low F-QC
interfacial free energy. Phase transitions can be observed from a kink in the
potential energy or density plotted against the variable we are changing (such
as the temperature or the pressure). However, we can further test for the nature
of the phase by calculating appropriate diffraction patterns: the
quasicrystal exhibits a distinctive 12-fold diffraction pattern
(Fig.~\ref{fig-quasicrystal}(f)), the Z phase exhibits a six-fold pattern, and
the fluid phase does not exhibit a well-defined pattern.

\begin{figure}
\begin{center}
\includegraphics{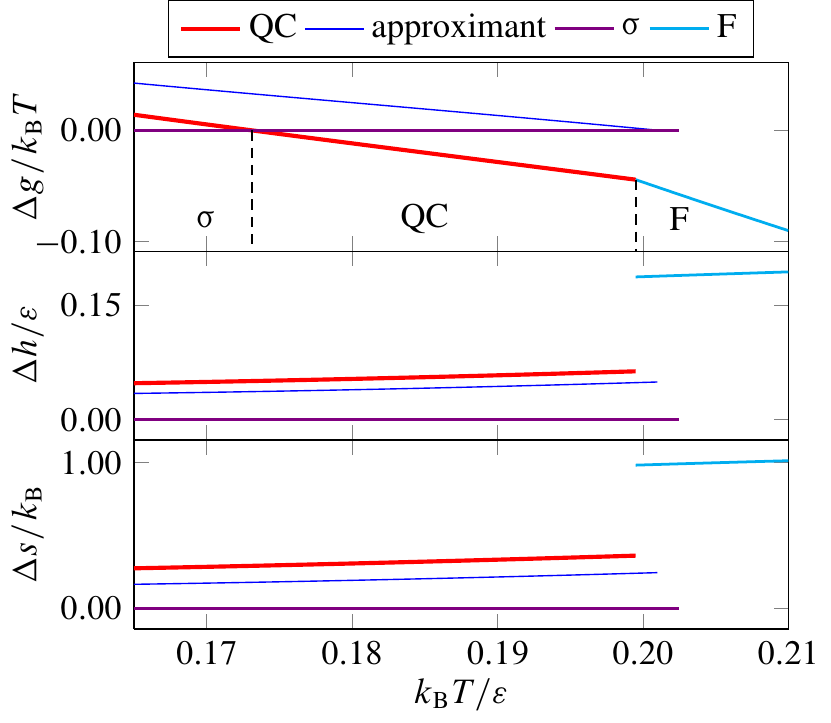}
\end{center}
\caption{Per-particle Gibbs energies, enthalpies and entropies,
relative to the $\upsigma$ phase, as a function of $T$ at $\sigma_\text{LJ}^2
p/k_\text{B}T =1.5$, $\sigma_\text{pw}=0.49$ and $n=6$. Lines end at the limit
of (meta)stability of the phases. Dashed lines denote coexistence points,
$k_\text{B}T_{\upsigma\leftrightarrow\text{QC}}/\varepsilon=0.173$ and
$k_\text{B}T_{\text{QC}\leftrightarrow\text{F}}/\varepsilon=0.1995$. The stable
phase in each region is explicitly marked. The error in $\upDelta g/k_\text{B}T$
is $<0.012\,k_\text{B}T$ \cite{Note1}.
}\label{fig-quasi-free-energies-BP1.5}
\end{figure}

Once a F-QC coexistence point is known, we can equate the chemical potential of
the fluid (as calculated by thermodynamic integration) with that of the
quasicrystal. From here, we can use thermodynamic integration to calculate the
chemical potential of the quasicrystal at other temperatures and pressures. By
also performing thermodynamic integration to find the chemical potential of the
$\upsigma$ phase, we can locate the putative $\upsigma$-QC transition. We plot
in Fig.~\ref{fig-quasi-free-energies-BP1.5} the free energies of the $\upsigma$,
QC and F phases as a function of $T$ at a constant $p/k_\text{B}T$. From this
figure, we can identify two clear transitions ($\upsigma$-QC and QC-F), and a
temperature window in which the quasicrystal is thermodynamically stable. In the
same figure, we also show the free energy of an approximant crystal phase that
is based on the most common packing of dodecagons in the quasicrystal; it is
never thermodynamically stable. To rationalise why the quasicrystal is stable,
we also calculate the enthalpy (obtained directly from simulations) and hence
obtain the entropy as a function of the temperature. These functions are also
depicted in Fig.~\ref{fig-quasi-free-energies-BP1.5}, and it is clear that in
the region in which the quasicrystal is stable, it does indeed have a higher
enthalpy than the $\upsigma$ phase, but its significantly higher entropy
nevertheless allows it to become more stable. The origin of this entropy is a
combination of the differing vibrational properties and the configurational
disorder present in the quasicrystalline phase. If we assume the vibrational
properties of the quasicrystal and its approximant are identical, the
configurational component of this entropy can be estimated from the entropy
difference between the quasicrystal and its approximant, $\upDelta s/k_\text{B}
\approx 0.113$ at $T_{\upsigma \leftrightarrow \text{QC}}$ \footnote{This value
of $\upDelta s$ is significantly larger than the tiling entropy of a random
Stampfli tiling, but somewhat smaller than that of a maximally random
square-triangle tiling, where both have been evaluated at zero phason strain
\cite{Oxborrow1993}.}.

From Fig.~\ref{fig-quasi-free-energies-BP1.5}, we can determine two coexistence
points on the phase diagram. We construct the remainder of the phase diagram
partly through the same procedure as above at, say, other pressures, but also
through the use of Gibbs--Duhem integration \cite{Kofke1993,*Kofke1993b,
Vega2008}. This method allows us to integrate Clapeyron equation analogues
to obtain new coexistence points from already known coexistence points,
but requires metastability and hysteresis in the transition so that the
simulated phases retain their identity when simulated at and around the
coexistence point. As a result, the method cannot be applied to the QC-F and
QC-Z transitions for much of the phase diagram due to their relative
reversibility.

\begin{figure}[tb]
\begin{center}
\includegraphics{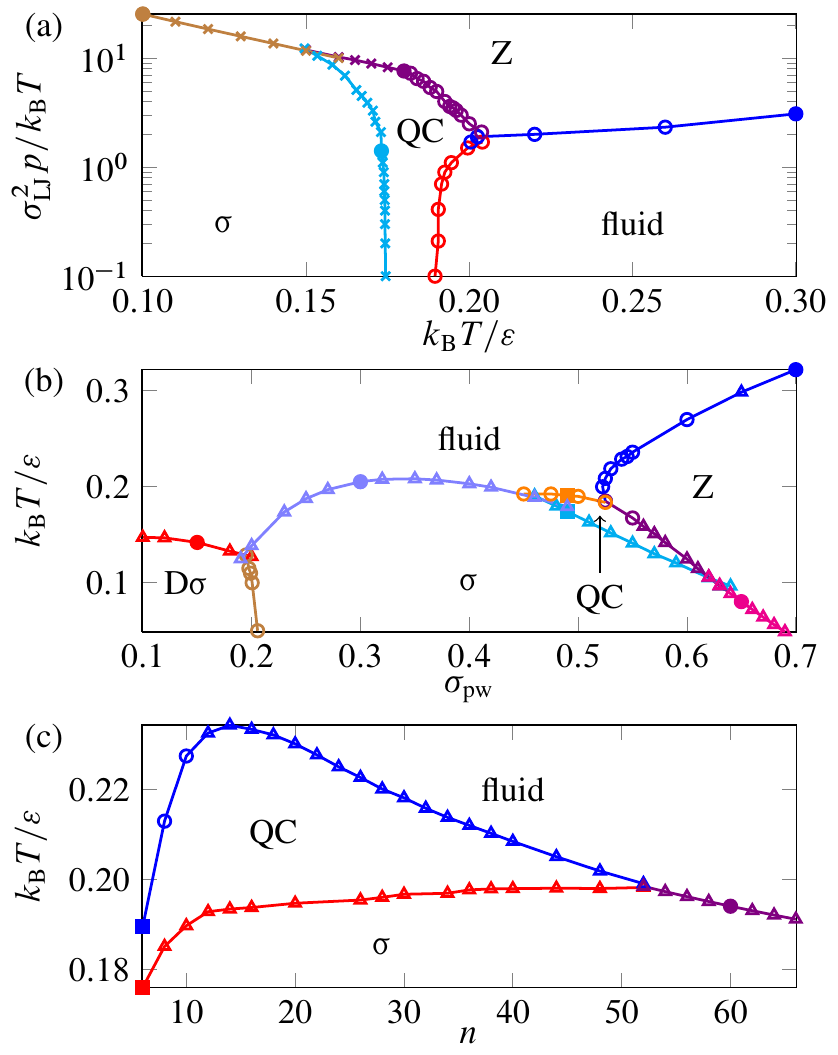}
\end{center}
\caption{Phase diagrams. Markers show the technique used to
obtain each point: Frenkel--Ladd and thermodynamic integration (filled circles),
Gibbs--Duhem integration (crosses), hamiltonian Gibbs--Duhem integration
(triangles) and direct simulation (open circles). Points obtained from previous
phase diagrams are shown as squares. (a) $p/T$-$T$ phase diagram;
$\sigma_\text{pw}=0.49$, $n=6$. (b) $T$-$\sigma_\text{pw}$ phase diagram;
$\sigma_\text{LJ}^2 p/k_\text{B}T =0.5$, $n=6$. (c) $T$-$n$ phase diagram;
$\sigma_\text{LJ}^2 p/k_\text{B}T=0.5$, $\sigma_\text{pw}=0.49$.}
\label{fig-quasicrystal-phasediags-all}
\end{figure}

The resulting phase diagram is shown in
Fig.~\ref{fig-quasicrystal-phasediags-all}(a). We note that the quasicrystal is
stable for a wide range of pressures. At very high pressures, the Z phase
becomes stable due to its greater density. The Z phase is a plastic crystal,
meaning that particles are able to rotate relatively freely: this is
understandable, as there is no obvious preferred way to orientate a five-fold
particle in a six-fold environment \footnote{We have also considered the
additional non-plastic hexagonal crystal structures suggested by Doppelbauer
\textit{et al.}\ for pentavalent patchy particles \cite{Doppelbauer2010}, but
these were found quickly to lose their rotational specificity at the
temperatures used in our simulations.}. No orientational ordering ensues in the
Z phase even at low temperatures. Whether it would form an
orientationally-ordered crystal or an orientational glass at sufficiently low
$T$ is not clear. It is noteworthy that for the QC-Z transition,
$\mathrm{d}p/\mathrm{d}T<0$; since $\rho_\text{Z} > \rho_\text{QC}$, the
Clapeyron equation implies that $S_\text{Z} > S_\text{QC}$: this is likely to
stem from the orientational entropy of the plastic crystal. Interestingly, in
addition to the two triple points ($\upsigma$-QC-Z and QC-Z-F), there is a range
of pressures where the stable thermodynamic phase changes from F to Z to QC to
$\upsigma$ as the system is cooled.

We also look at the effect of modifying potential parameters on the phase
diagram, thus allowing us to assess the extent of quasicrystal stability and how
finely-tuned the particle properties would need to be to observe a
quasicrystalline phase in experiment. To determine the phase diagram as the
potential parameters change, we additionally use hamiltonian Gibbs--Duhem
integration, whereby we numerically integrate a Clapeyron equation that has been
generalised to allow for changes to the potential itself \cite{Vega2008}.

We first investigate the behaviour of the system as the patch width is varied.
From the phase diagram in Fig.~\ref{fig-quasicrystal-phasediags-all}(b), we see
that the quasicrystal is only stable for a limited range of $\sigma_\text{pw}$,
although the range may perhaps be wider at other pressures. As the patch width
narrows, the quasicrystal becomes increasingly enthalpically destabilised with
respect to the $\upsigma$ phase, as the many six-fold environments cannot
satisfactorily fulfil patch-patch interactions. By contrast, as the patch width
increases, the quasicrystal becomes more stable with respect to the $\upsigma$
phase, but the hexagonal phase is stabilised even more: as the patches are wider
and closer to the isotropic case, six-fold environments are enthalpically
preferred.

Another noteworthy feature of this phase diagram is that below
$\sigma_\text{pw}\approx 0.35$, $T_{\upsigma\leftrightarrow\text{F}}$ begins to
decrease: as the patches become narrower, there is an increasing energetic
penalty for patches that do not point directly at each other. For very narrow
patches, the $\upsigma$ phase transforms into a distorted $\upsigma$ phase
(D$\upsigma$) \cite{Doppelbauer2010}, in which three of the five patches point
directly at the patches of neighbouring particles, thus breaking the square
symmetry of the lattice. The reason for this behaviour is that at narrow patch
widths, only `perfect' connections count, as a slight misalignment of the
patches rapidly reduces the interparticle attraction, and three perfect
interactions become favoured over five imperfect ones.

Because colloidal interactions are often quite short-ranged, we also wish to
investigate what happens to the system when we change the LJ
exponent ($n$), which changes the range of the potential and the width of the
potential well. We show the phase diagram as a function of $n$ in
Fig.~\ref{fig-quasicrystal-phasediags-all}(c).
The effects of increasing $n$ on isotropic potentials are well understood
\cite{Hagen1994, *Vliegenthart1999}: the liquid phase is energetically
destabilised because the intrinsic disorder in interparticle neighbour distances
is penalised by the narrower potential \cite{Doye1996}. In
Fig.~\ref{fig-quasicrystal-phasediags-all}(c), the range of stability of the
quasicrystal initially increases. The enthalpies of the QC and $\upsigma$ phases
increase only slightly as $n$ increases, because their relative order means that
most interparticle neighbour distances can be close to optimal, but do so more
for the quasicrystal, as some disorder is typically present in the quasicrystal
configurations (for example, particles with unclassified environments in
Fig.~\ref{fig-quasicrystal}(d)); the $\upsigma$-QC coexistence temperature does
therefore increase slightly. By contrast, the fluid is initially much more
enthalpically penalised: the fluid enthalpy increases rapidly, and the fluid
density decreases as $n$ increases. At the maximum in
$T_{\text{QC}\leftrightarrow\text{F}}$ ($n\approx 14$), the size of the
temperature window of quasicrystal stability is three times that of the $n=6$ LJ
potential. However, a narrower potential well associated with increasing $n$
also means that there is less vibrational entropy associated with the ordered
structures, and beyond $n\approx 14$, this effect becomes dominant. The QC-F
coexistence temperature decreases until the quasicrystal loses stability at
$n\approx 53$.

The potential at the higher values of $n$ shown in
Fig.~\ref{fig-quasicrystal-phasediags-all}(c) is very short-ranged indeed. The
fact that the quasicrystal maintains and even increases its window of stability
for ranges of attraction typical of colloidal particles suggests that this is a
robust, general result and that stable quasicrystals can be expected to form in
experimental realisations of this system.

There are several potential approaches to such experimental realisations. If
patchy colloidal particles similar to the ones studied here could be created, it
would be reasonable to expect that a quasicrystal may form when they are
confined in two dimensions. For example, Chen \textit{et al.}\ observed the
formation of a 2D kagom\'e lattice from tri-block Janus particles by introducing
a density mismatch with the solvent to confine their colloidal system into two
dimensions \cite{Chen2011}.  A possible alternative to using colloidal patchy
particles might involve the use of DNA multi-arm motifs \cite{Yan2003, *He2005b,
*He2006, *Zhang2008}, for which a variety of 2D crystalline arrays have
been observed for motifs with different numbers of arms, including a $\upsigma$
phase for five-arm motifs that is analogous to what we see in the current model
when patches are narrow. However, such DNA motifs have a well-defined `valence',
and there is no equivalent of the patch width that could be tuned to make two
co-ordination numbers compete. Therefore, a possible approach to forming a DNA
quasicrystal might be to use a two-component mixture of five-arm and six-arm
motifs of the appropriate composition.

In summary, we have shown how we can compute if and where a quasicrystalline
phase is thermodynamically stable. To the best of our knowledge, this is the
first such calculation of the chemical potential of a quasicrystal and the
associated phase diagrams that has been obtained directly from simulations with
no approximations. For our patchy particle system, we found that the
quasicrystalline phase is stable over a significant portion of the phase diagram
and is stabilised primarily by its configurational entropy. It is robust to
parameter changes in the model, which inspires confidence that such a
thermodynamically stable quasicrystal might be experimentally realised.

\begin{acknowledgments}
We wish to thank the Engineering and Physical Sciences Research Council for
financial support.
\end{acknowledgments}

%


\vspace{2cm}

\appendix
\noindent{\large \textbf{Supplementary material}}

\section{Simulation details}

The simulations described in our work used the Metropolis Monte Carlo approach \cite{Metropolis1953} in the isobaric-isothermal ensemble \cite{Frenkel2002,Eppenga1984} with periodic boundary conditions.

The potential as discussed in the manuscript is given more formally by
\begin{equation}
V(\mathbold{r}_{ij},\varphi_i,\varphi_j)=\begin{cases}
V^{\text{LJ}}(r_{ij}) & r_{ij}<\sigma_\text{LJ}, \\
V^{\text{LJ}}(r_{ij})  V^{\text{A}}(\hat{\mathbold{r}}_{ij},\varphi_i,\varphi_j) &  \sigma_\text{LJ}\le r_{ij} , \\
\end{cases}\label{quasi-eqn-potential}
\end{equation}
where $\mathbold{r}_{ij}$ is the interparticle vector connecting the centres of the two particles $i$ and $j$, $r_{ij}$ is its magnitude,  $\varphi_i$ and $\varphi_j$ are the orientations of the particles $i$ and $j$, the generalised Lennard-Jones potential is
\begin{equation}
V^{\text{LJ}}(r_{ij}) =
4 \varepsilon \left[\left(\sigma_\text{LJ}/r_{ij}\right)^{2n}-\left(\sigma_\text{LJ}/r_{ij}\right)^{n}\right],\label{quasi-eqn-LJ}
\end{equation}
where $n=6$ for the standard Lennard-Jones potential, and
\begin{equation}
V^{\text{A}}(\hat{\mathbold{r}}_{ij},\,\varphi_i,\,\varphi_j) = \exp\left[\frac{-\theta_{k_\text{min}ij}^2}{2\,\sigma_\text{pw}^2}\right]\exp\left[\frac{-\theta_{l_\text{min}ji}^2}{2\,\sigma_\text{pw}^2}\right],
\end{equation}
where $\sigma_\text{pw}$ is the patch width, $\theta_{kij}$ is the angle between the patch vector of patch $k$ on particle $i$ and the interparticle vector $\mathbold{r}_{ij}$, and $k_\text{min}$ is the patch that minimises the magnitude of this angle. Two particles therefore interact only through a single pair of patches.
In the simulations reported here, we use a potential cutoff of $r_\text{cut}=3\sigma_\text{LJ}$, and the potential in Eq.~\eqref{quasi-eqn-LJ} is shifted so that it equals zero at $r_{ij}=r_\text{cut}$. The crossover to including angular modulation in the potential (Eq.~\eqref{quasi-eqn-potential}) is likewise adjusted so that it still occurs when the potential energy is identically zero.

Most simulations reported used relatively small system sizes (of the order of 500 particles), but the phase diagram in Fig.~\ref{fig-quasicrystal-phasediags-all}(a) and the free energies in Fig.~\ref{fig-quasi-free-energies-BP1.5} were also calculated using a system of the order of 2500 particles, and certain transitions were calculated with systems of 5000 and 10000 particles, with no detectable differences in the behaviour of the systems, confirming that reasonably small simulations were sufficient to calculate phase diagrams.

\section{Summary of methods}
Here, we summarise the methods we used in the manuscript in somewhat more detail. The approaches described here are all standard and are provided here as a quick summary for the benefit of the reader only.

\subsection{Direct coexistence \textit{vs} brute force simulation}
Direct coexistence simulations \cite{Ladd1977, Vega2008} are simulations in which an explicit interface is set up between the two phases in question, and the thermodynamically stable phase is expected to grow into the thermodynamically unstable phase such that the interface between the two phases moves. In brute force simulations, no interface is set up; instead, a single phase is simulated and it spontaneously converts into the thermodynamically stable phase. Typically, this brute force simulation process is prone to hysteresis effects, where for example the quasicrystal will form at some temperature $T_1$, but when the quasicrystal is heated in a reverse brute force simulation, it will melt at temperature $T_2$, where $T_2>T_1$, because nucleation involves a free energy barrier that must be overcome. In direct coexistence simulations, this hysteresis does not occur, because both phases are present from the start, and the interface between them is pre-formed.

In our simulations, we established that there is essentially no hysteresis between the quasicrystal and the fluid under the conditions described in this paragraph, and so direct coexistence simulations offer no advantage over brute force simulations. Indeed, because nucleation is so facile, numerous phase interfaces are spontaneously formed in a direct coexistence simulation, and so direct coexistence simulations were not useful in phase diagram determination at this stage.

However, we have used direct coexistence simulations in certain cases as a check in confirming that the Frenkel--Ladd simulations yielded the correct coexistence points.

\subsection{Thermodynamic integration}
Thermodynamic integration \cite{Frenkel1984, Vega2008} refers to a series of integration schemes of thermodynamic potentials.
For integration along iso-($\beta p$) curves, where $\beta \equiv 1/k_\text{B} T$, we start from the product rule
\begin{equation}
\pdc{(G/T)}{T}{(\beta p)} = \frac{1}{T} \pdc{G}{T}{(\beta p)} - \frac{G}{T^2},\label{eqn-gibbsHelm-bp}
\end{equation}
from where an application of simple thermodynamics allows us to write
\begin{equation}
 \pdc{(G/T)}{T}{(\beta p)} = -\frac{U}{T^2}.
\end{equation}
This can be integrated to give
\begin{equation}
 \frac{G_2}{N k_\text{B} T_2} = \frac{G_1}{N k_\text{B} T_1} - \int_{T_1}^{T_2} \frac{U}{N k_\text{B} T^2}\,\der T,
\end{equation}
and this is the result we require for thermodynamic integration. Provided we know the Gibbs energy at some $T_1$, then we can find the Gibbs energy at $T_2$ along the iso-($\beta p$) curve by running several $NpT$ simulations for temperatures between $T_1$ and $T_2$ (at fixed $\beta p$), and finding a polynomial fit to the integrand, which can then by integrated analytically. Analogous results arise for integration along isobars \cite{Vega2008},
\begin{equation}
 \frac{G_2}{N k_\text{B} T_2} = \frac{G_1}{N k_\text{B}  T_1} - \int_{T_1}^{T_2} \frac{H}{N k_\text{B} T^2}\,\der T,
\end{equation}
and isotherms \cite{Vega2008},
\begin{equation}
 \frac{A_2}{N k_\text{B} T} = \frac{A_1}{N k_\text{B} T} +  \int_{\rho_1}^{\rho_2} \frac{p}{k_\text{B} T\rho^2}\,\der \rho.
\end{equation}

It is also possible to extend the scheme by coupling to a different hamiltonian, such that the potential energy is given by $U=U_\text{real} + \lambda U_\text{extra}$, and then integrating along $\lambda$; we find that \cite{Vega2008}
\begin{equation}
A(\lambda=1) = A(\lambda=0) + \int_{0}^1 \avg{U_\text{extra}}\,\der \lambda.
\end{equation}
This is known as hamiltonian thermodynamic integration \cite{Vega2008} and is one component of the Einstein crystal Frenkel--Ladd approach \cite{Frenkel1984}.

The free energy of a fluid phase evaluated from the ideal gas by decreasing the pressure at constant temperature is given by \cite{Vega2008}
\begin{equation}
 \frac{A}{Nk_\text{B}T}=\ln(\rho \Lambda^3) - 1  + \int_0^\rho \left[ \frac{p}{k_\text{B} T\rho^2} - \frac{1}{\rho} \right] \,\der \rho,
\end{equation}
where $\Lambda$ is the de Broglie thermal wavelength. It is usually beneficial to fit the entire integrand to a single polynomial to avoid diverging terms. As the density tends to zero, the integrand tends to the second virial coefficient \cite{Vega2008}; it is helpful to evaluate this explicitly to benchmark the simulation results.

\subsection{Gibbs--Duhem integration}
In Gibbs--Duhem integration \cite{Kofke1993,*Kofke1993b, Vega2008}, we numerically integrate the Clapeyron equation,
\begin{equation}
 \deriv{p}{T} = \frac{\upDelta H}{T\upDelta V},
\end{equation}
or an analogous expression, in order to obtain new coexistence points from already known coexistence points. In our simulations, we used the fourth-order Runge--Kutta algorithm to perform this integration.

\subsection{Frenkel--Ladd Einstein crystal approach}
The Einstein crystal approach \cite{Frenkel1984, Vega2008} is a fairly involved way of calculating the free energy of a crystal from the Einstein crystal, for which the free energy is known. We direct the interested reader to Frenkel and Smit's book (Ref.~\cite{Frenkel2002}) and the review paper by Vega and co-workers (Ref.~\cite{Vega2008}) for an in-depth discussion of the method. In the Frenkel--Ladd scheme as implemented for our model, we take the rotational energy to be
\begin{equation}
u_{\text{Ein, or},\,i} = \Lambda_\text{rot} \sin^2 \left[\frac{p}{2} \left(\varphi_i - \varphi_{i,\,\text{orig}} \right)\right],
\end{equation}
where $\Lambda_\text{rot}$ is a constant that we vary and $p=5$ is the number of patches. We thus ensure that the particle's symmetry is accounted for: rotations into degenerate positions give the same Frenkel--Ladd energy.

\section{Reliability of free energy calculations}
Computing absolute free energies from simulations requires a combination of
numerical techniques, each of which can introduce errors, both systematic
and random. Since it is difficult to keep track of how the errors propagate, the
approach which is usually undertaken is to perform \textit{a posteriori} checks
on the whole process. As discussed by Vega and co-workers \cite{Vega2008}, `consistency checks' can be applied to simulations to ensure that the data obtained in free energy calculations are self-consistent and correspond to reality. There are several such checks that can be performed. For example, we can calculate the free energy at a particular point using several routes (\textit{e.g.}~by integrating along different isotherms, isobars or iso-($\beta p$) curves to the same point). Furthermore, direct coexistence simulations can be used to ensure that the transition temperature does in fact occur where free energy curves intersect. Finally, in simulations involving Gibbs--Duhem integration, we can integrate first in the `forward' and then in the `reverse' direction, ensuring that the method reproduces the starting point. We have performed all these checks to ensure that the results we have obtained are robust.

Gao and co-workers compared the results of Debye crystal reference state calculations with different force constants to obtain an estimate of the error in the free energy of hexagonal ice \cite{Gao2000}. We do something similar to estimate the error in free energy in Fig.~\ref{fig-quasi-free-energies-BP1.5} in the main paper: we calculate the Frenkel--Ladd free energy at several points ($k_\text{B}T/\varepsilon=0.1$, $k_\text{B}T/\varepsilon=0.15$, $k_\text{B}T/\varepsilon=0.17$ and $k_\text{B}T/\varepsilon=0.19$) along the iso-($\beta p$) curve, and then use thermodynamic integration to obtain free energies from each of the starting Frenkel--Ladd free energies. The largest difference in free energy between any pair of calculations along the entire curve can then serve as an estimate of the error in free energy in our simulations. The \textit{largest} such deviation was $\upDelta (\upDelta g)/k_\text{B}T = 0.012$, which is sufficiently small not to affect any conclusion, and the error in practice is likely to be significantly smaller, as we were able to confirm phase transition temperatures in independent direct coexistence simulations.

\end{document}